\documentclass[a4paper,11pt]{article}
\usepackage{pos}

\usepackage{booktabs,array}
\usepackage[utf8]{inputenc}
\usepackage{float,graphicx,hhline}
\usepackage{algorithm,algpseudocode}
\usepackage{appendix}
\usepackage{mathtools}
\usepackage{multirow}
\usepackage{tikz-cd}
\usepackage{slashed}
\usepackage{soul}
\usepackage{pifont}
\usepackage{tikzducks}
\usepackage[caption=false]{subfig}
\usepackage[export]{adjustbox}
\usepackage{enumitem} 
\usepackage{bbm}
\usepackage{dsfont}
\usepackage{hyphenat}
\usepackage{comment}
\AtBeginDocument{
\heavyrulewidth=.08em
\lightrulewidth=.05em
\cmidrulewidth=.03em
\belowrulesep=.65ex
\belowbottomsep=0pt
\aboverulesep=.4ex
\abovetopsep=0pt
\cmidrulesep=\doublerulesep
\cmidrulekern=.5em
\defaultaddspace=.5em
}

\usepackage{hyperref}
\hypersetup{
    pdftitle={},
    colorlinks=true,     
    linkcolor=blue,      
    citecolor=blue,      
    filecolor=blue,      
    urlcolor=blue        
}
\usepackage{cleveref}
\Crefname{equation}{Eq.}{Eqs.}

\DeclareMathAlphabet{\mathbbold}{U}{bbold}{m}{n}

\let\Re\undefined
\DeclareMathOperator{\Re}{Re}
\DeclareMathOperator{\IM}{Im}
\DeclareMathOperator{\Tr}{Tr}

\title{Sampling QCD field configurations with gauge-equivariant flow models }

\author[a,b]{Ryan~Abbott}
\author[c]{Michael~S.~Albergo}
\author[g]{Aleksandar~Botev}
\author[a,b,d]{Denis~Boyda}
\author[c,e]{Kyle~Cranmer}
\author[a,b]{Daniel~C.~Hackett}
\author[a,b,f]{Gurtej~Kanwar}
\author[g]{Alexander~G. D. G.~Matthews}
\author[g]{S\'{e}bastien~Racani\`{e}re}
\author[g]{Ali~Razavi}
\author[g]{Danilo~J.~Rezende}
\author[a,b]{Fernando~Romero-L\'opez}
\author*[a,b]{Phiala~E.~Shanahan}
\author[a,b,h]{Julian~M.~Urban}

\affiliation[a]{Center for Theoretical Physics, Massachusetts Institute of Technology, Cambridge, MA 02139, USA}
\affiliation[b]{The NSF AI Institute for Artificial Intelligence and Fundamental Interactions}
\affiliation[c]{Center for Cosmology and Particle Physics, New York University, New York, NY 10003, USA}
\affiliation[d]{Argonne Leadership Computing Facility, Argonne National Laboratory, Lemont IL 60439, USA}
\affiliation[e]{Physics Department, University of Wisconsin-Madison, Madison, WI 53706, USA.}
\affiliation[f]{Albert Einstein Center, Institute for Theoretical Physics, University of Bern, 3012 Bern, Switzerland}
\affiliation[g]{DeepMind, London, UK}
\affiliation[h]{Institut f\"ur Theoretische Physik, Universit\"at Heidelberg, Philosophenweg 16, 69120 Heidelberg, Germany}

\emailAdd{phiala@mit.edu}

\abstract{
Machine learning methods based on normalizing flows have been shown to address important challenges, such as critical slowing-down and topological freezing, in the sampling of gauge field configurations in simple lattice field theories. A critical question is whether this success will translate to studies of QCD. This Proceedings presents a status update on advances in this area. In particular, it is illustrated how recently developed algorithmic components may be combined to construct flow-based sampling algorithms for QCD in four dimensions. The prospects and challenges for future use of this approach in at-scale applications are summarized.}

\FullConference{%
 The 39th International Symposium on Lattice Field Theory, LATTICE2022 8th-13th August 2022, Bonn Germany
}

\begin{document}

\maketitle

\section{Introduction}
The rapid development of machine learning, and in particular deep learning, over the last decade has spawned a new wave of algorithms for computational physics~\cite{Carrasquilla:2020mas,Boehnlein:2021eym,Dawid:2022fga}.
In lattice quantum field theory, custom machine learning tools are being developed to accelerate almost every step of the computational workflow~\cite{Boyda:2022nmh}, from configuration generation~\cite{Nicoli:2020njz,medvidovic2021generative,Foreman:2021ixr,Foreman:2021ljl,Finkenrath:2022ogg, Wang2017,Huang:2017,song2017nice,Tanaka:2017niz,levy2018generalizing,Pawlowski:2018qxs,Cossu:2018pxj,Wu:2019,Bachtis:2020dmf,Nagai:2020jar,Tomiya:2021ywc,Bachtis:2021eww,Wu:2021tfb,Gerdes:2022eve,Alexandru:2017czx,Lawrence:2021izu,Wynen:2020uzx,LiWang2018NNRG,Albergo:2019eim,Nicoli:2020evf,Kanwar:2020xzo,Boyda:2020hsi,Albergo:2021bna,Hackett:2021idh,Gabrie:2021tlu,DelDebbio:2021qwf,Foreman:2021rhs,Jin:2022bgq,deHaan:2021erb,Singha:2022lpi,Matthews:2022sds,Caselle:2022acb,Pawlowski:2022rdn,Albergo:2022qfi,Albergo:2021vyo,Favoni:2020reg,Abbott:2022zhs} to calculation or design of observables~\cite{Nicoli:2020njz,Boyda:2020nfh,Bachtis:2020fly,Bachtis:2020dmf,Palermo:2021jrf,Tan:2021cgs,Li:2017xaz,Wetzel:2017ooo,Alexandrou:2019hgt,Blucher:2020mjt,Yau:2020emg,Zhou:2018ill,Detmold:2021ulb,hu2017discovering,wetzel2020discovering} and various aspects of analysis~\cite{Yoon:2018krb,Zhang:2019qiq,Shanahan:2018vcv,Hudspith:2021iqu,Kades:2019wtd,Offler:2019eij,Horak:2021syv,Chen:2021giw,Wang:2021cqw,Shi:2022yqw,DelDebbio:2020rgv,Karpie:2019eiq}, while maintaining rigorous guarantees of exactness.

One particular approach to accelerating gauge field generation via machine learning is based on the use of flow transformations as trivializing maps~\cite{Luscher:2009eq,rezende2016variational,dinh2017density,JMLR:v22:19-1028}, as outlined in Sec.~\ref{sec:flowtransforms}. Bespoke flow architectures tailored to this application have been constructed, including gauge-equivariant architectures designed for Abelian and non-Abelian gauge field theories~\cite{Kanwar:2020xzo,Boyda:2020hsi}, and methods to incorporate fermions~\cite{Albergo:2021bna,Abbott:2022zhs}. Proof-of-principle implementations have demonstrated success in ameliorating important challenges such as critical slowing-down and topological freezing, as well as the computation of thermodynamic quantities, in theories ranging from scalar field theory~\cite{Albergo:2019eim,Nicoli:2020njz,Nicoli:2021inv,DelDebbio:2021qwf,Singha:2022lpi} through Yukawa theory~\cite{Albergo:2021bna}, to the Schwinger model~\cite{Albergo:2022qfi,Finkenrath:2022ogg} and SU(3) gauge field theory with fermions in two dimensions~\cite{Abbott:2022zhs}.

The following sections outline how recent developments may be combined to construct flow transformations to enable sampling of QCD gauge field configurations, and present the first numerical demonstration of this technique applied to lattice QCD.

\section{Flow transformations of lattice field configurations}
\label{sec:flowtransforms}

In the machine learning lexicon, a flow transformation~\cite{rezende2016variational,dinh2017density,JMLR:v22:19-1028} is a diffeomorphism between manifolds, defined with a large number of free parameters such that it can be optimized, or trained, to some objective. Precisely, a flow $f$ maps samples $z$ from a `prior' or base distribution $r(z)$ to samples $\varphi=f(z)$ distributed according to 
\begin{equation}
	q(\varphi) = r(z) \rvert \det \partial f(z)/ \partial z \rvert ^{-1} \ .
\end{equation}
The free parameters of $f$ are optimized such that $q(\varphi)$ approximates a `target' probability distribution $p(\varphi)$, i.e., $q(\varphi)\simeq p(\varphi)$. Typically this is achieved by minimization of a `loss function' that quantifies the difference between $q(\varphi)$ and $p(\varphi)$, such as the Kullback-Leibler (KL) divergence~\cite{Kullback:1951} or related measures. It is often convenient to define the loss function such that it can be computed stochastically using only samples from $r(z)$ (known as `self-training' as no `training data', i.e., no samples from $p(\varphi)$, are needed). Given a flow model $f$, samples from $p(\varphi)$ can be obtained by using samples drawn from the model distribution $q(\varphi)$ as proposals in the independence Metropolis algorithm~\cite{Metropolis:1953am,Hastings:1970aa,tierney1994markov}, or by reweighting.

Applications of flow models to lattice field theory are discussed in Refs.~\cite{LiWang2018NNRG,Albergo:2019eim,Kanwar:2020xzo,Boyda:2020hsi,Hackett:2021idh,Albergo:2021bna,Albergo:2021vyo,Albergo:2022qfi,Nicoli:2020evf,Nicoli:2020njz,Foreman:2021ixr,Foreman:2021ljl,Foreman:2021rhs,DelDebbio:2021qwf,Gabrie:2021tlu,deHaan:2021erb,Lawrence:2021izu,Jin:2022bgq,Pawlowski:2022rdn,Finkenrath:2022ogg,Gerdes:2022eve,Singha:2022lpi,Matthews:2022sds,Caselle:2022acb}. In the context of gauge field configuration generation, $\varphi$ (and $z$) are field configurations, and the goal is to sample efficiently from $p(\varphi)=\frac{1}{\mathcal Z} e^{-S(\varphi)}$, where $S(\varphi)$ is the Euclidean action of the theory and $\mathcal{Z}$ is a normalization constant. In the most direct approach, $r(z)$ can be taken as a trivial distribution (e.g., an independent uniform distribution over the Haar measure on each link for a gauge theory), and $f$ implements an approximation of a trivializing map~\cite{Luscher:2009eq}.
Naturally, many other choices are possible and may be practical for particular applications; for example, $r(z)$ may represent the distribution of configurations of a pure-gauge theory that can be sampled with the Heat Bath algorithm~\cite{Creutz:1979dw,Cabibbo:1982zn,Kennedy:1985nu}, or a theory with a different action or with different parameters from $p(\varphi)$. Alternatively, rather than using the flow directly for sampling, the entire construction may be embedded as a component of a more general or hybrid sampling algorithm~\cite{Foreman:2021ixr,Foreman:2021ljl,Finkenrath:2022ogg}. Variations of flow transformations may also be used in other applications in lattice field theory, such as contour deformation and density-of-states approaches to ameliorating the sign problem~\cite{Yamauchi:2021kpo,Lawrence:2022afv,Pawlowski:2022rdn,Detmold:2021ulb}.

Naturally, the efficiency of sampling gauge field configurations using a flow transformation, regardless of where the flow construction appears in the sampling algorithm, will depend both on the level of optimization that has been achieved and on the flexibility, or expressivity, of the function $f$ itself. Using architectures that respect important features of the target distribution, such as gauge symmetry~\cite{Boyda:2020hsi}, can be critical to achieving practical success.

\section{Numerical demonstration}
\label{sec:numericaldemo}

\begin{figure}[t!]
    \centering
    \includegraphics[width=0.7\linewidth]{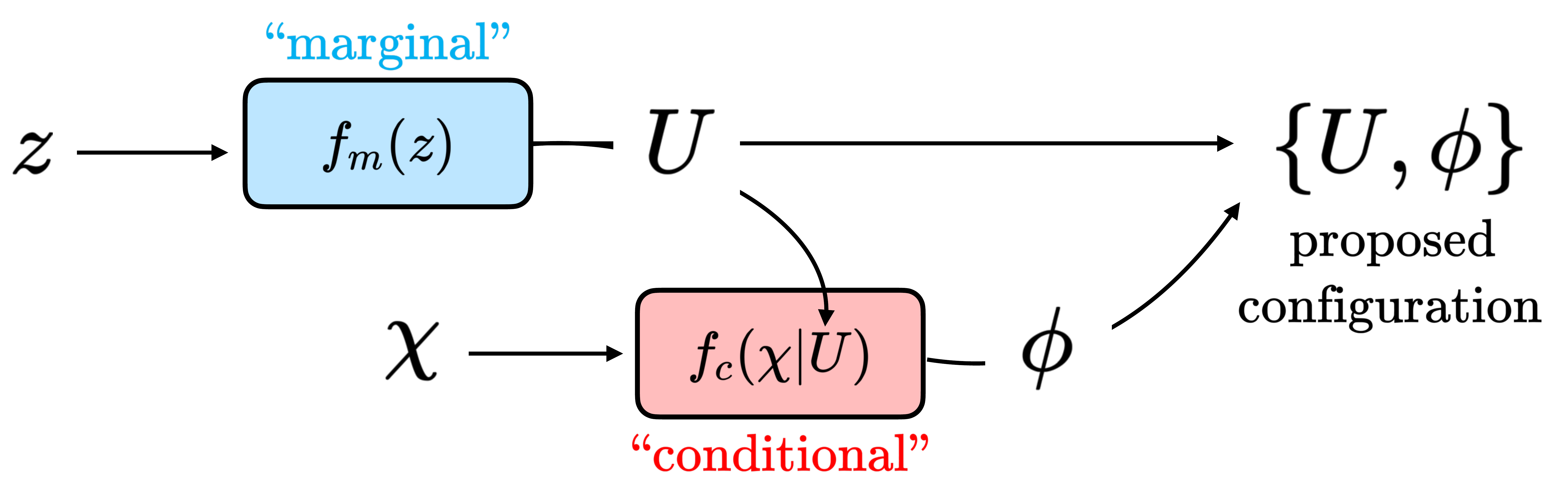}
    \caption{Sketch of the workflow of a joint model, reproduced from Ref.~\cite{Abbott:2022zhs}. The notation $z$ and $\chi$ denotes samples from the base distribution, transformed via the marginal and conditional models $f_m(z)$ and $f_c(z|U)$ to produce the gauge and pseudofermion fields, $U$ and $\phi$. The proposed configuration is distributed as ${q(\phi, U) = q(U) q(\phi |U)}$, in terms of the marginal and conditional model probability densities $q(U)$ and $q(\phi |U)$.}
    \label{fig:diag}
\end{figure}

Combining the developments of Refs.~\cite{Boyda:2020hsi,Abbott:2022zhs}, all of the necessary components are in place to apply flow-based sampling of lattice gauge field configurations to QCD. This section outlines the first demonstration of this application.

For illustration, an architecture is optimized for flow-based sampling of QCD with $N_f=2$ fermion flavors in 4D, with lattice volume $4^4$, strong coupling $\beta=1$, and $\kappa=0.1$. As in Ref.~\cite{Abbott:2022zhs}, the flow is composed of two separate flow models---marginal and conditional submodels---to produce samples of the gauge and pseudofermion degrees of freedom. The resulting joint model structure is illustrated in Fig.~\ref{fig:diag}. 
As detailed in Ref.~\cite{Abbott:2022zhs}, an efficient way to use joint models is to draw multiple pseudofermion samples for fixed gauge fields and average the resulting weight factors for the Metropolis accept-reject or reweighting procedure. This yields more precise estimators of the determinant of the Dirac operator; in the limit of an infinite number of pseudofermion samples, this would give the same results as 
using the marginal flow model with an exact evaluation of the determinant of the Dirac operator. 

\subsection{Model definition}

{\it \noindent Marginal model}

The marginal model uses a Haar-uniform prior distribution and consists of 48 gauge-equivariant spline coupling layers~\cite{Boyda:2020hsi}, with convolutional neural networks acting in the spectral flow to define the spline parameters. Each spectral flow is parametrized by two convolutional neural networks, each made of four convolutional layers with kernel size 3. The two networks are structurally identical except for the number of input channels, which is determined by the number of plaquettes and concatenated input (see~\cite{Boyda:2020hsi} for details). The two intermediate convolutions each have 32 input and output channels. The last layer has 32 input channels and 49 output channels to parametrize 16 spline knots and an offset. The ``Leaky~ReLU'' activation function was used in intermediate layers, but no activation is applied to the final output. Spatially separated convolutions are used; i.e., the four-dimensional convolutional neural networks consist of blocks, where every block is the sequential application of a one-dimensional convolution along each of the four dimensions, with a bias for only the last one. The number of channels between each one-dimensional convolution is set to the number of output channels. The weights and biases of the convolutions comprise all free parameters of the architecture. The gauge-invariant inputs to the inner spectral flow are the real and imaginary parts of the traces of all frozen plaquettes. Masking patterns are generated with the algorithm for higher-dimensional masks presented in Ref.~\cite{higherdim}, using the parameters \textit{spatial-dims} = [ 4, 4, 4, 4 ], \textit{mask-orientations} = [ 0 ], \textit{width} = 4, \textit{active-phase} = 0, \textit{orientations-to-shift} = [ 0, 1, 2, 3 ], and \textit{shifts} = [ 1, 2, 1, 1 ]. These parameters are varied across flow layers using the alternation scheme that always sets \textit{mask-orientations}[0] to \textit{orientations-to-shift}[0], also described in Ref.~\cite{higherdim}.

The marginal model is constructed and trained prior to the conditional model being trained.
The convolution weights are first initialized using the Xavier normal~\cite{He:2015dtg} scheme with gain parameter $0.5$, and biases are set to zero.
After initialization, self-training proceeds by minimizing stochastic estimates of the ``reverse'' KL loss~\cite{Albergo:2019eim} computed with flow model samples, as discussed above.
Gradients of the fermion determinant are estimated stochastically as described in Appendix~C of Ref.~\cite{Albergo:2021bna}.
During training, the Dirac operator is regulated as $(D D^\dagger + \mu_0)$, with $\mu_0 = 10^{-5}$; no regulator is used in evaluation. The marginal model is trained for 21k steps with batch size 512. The initial learning rate is $10^{-3}$, and decreases by a factor 0.8 every 10k steps. All optimization is performed with the Adam optimizer~\cite{kingma2017adam} with $\epsilon= 10^{-2}$. Clipping is applied to the norm and value of gradients~\cite{zhang2020gradient}; the maximum norm value was set to 10, and maximum value to 0.1. \\ 

{\it \noindent Conditional model}

The conditional model uses a Gaussian prior and consists of 36 pseudofermion layers built from parallel transport convolutional networks (PTCN), defined in Ref.~\cite{Abbott:2022zhs}. The masking pattern rotates through the spin projectors $\frac{1}{2} (1 \pm \gamma_{\mu})$ for $\mu \in \{0,1,2,3,5\}$ along with even/odd spatial projectors. In the notation of Ref.~\cite{Abbott:2022zhs}, the PTCNs use $n_{PT} = 6$ and $H = 4$. The context function of each PTCN is a convolutional neural network---as described in the Appendix B.2 of Ref.~\cite{Abbott:2022zhs}---and it has six input channels, namely $\sum_{\mu \neq \nu}$, $\Re \Tr P_{\mu\nu}$, $\sum_{\mu \neq \nu} \IM \Tr P_{\mu\nu}$, $I_0$, $I_1$, $I_2$, and $I_3$ where $P_{\mu\nu}(x)$ is the plaquette and $I_i(x) = x_i \, \mathrm{mod} \, 4$. The part of the context function that is shared among all PTCNs has 8 hidden channels, with ``ELU'' used as the inner activation function. The final activation function is ``tanh''. All layers have biases and use kernel size 3.

The weights in the networks defining the conditional model architecture are initialized with the Xavier normal~\cite{He:2015dtg} scheme with a gain of $0.3$. The biases are initialized using a Gaussian distribution with a mean of 0 and a standard deviation of 0.05.
Conditional model training then proceeds by self-training the joint model for gauge and pseudofermion fields, similar to Ref.~\cite{Abbott:2022zhs}, except here the marginal model is kept fixed.
The conditional model is optimized in 3 phases: for 16k steps with batch size 640, learning rate $10^{-4}$; for 28k steps with batch size 640, learning rate $5 \times 10^{-5}$, stepped to $2.5 \times 10^{-5}$ after 20k steps; for 17k steps with batch size 800, learning rate $2.5 \times 10^{-5}$. During training, the Dirac operator is regulated as described above for the marginal model. The Adam optimizer~\cite{kingma2017adam} is used with the same settings as for the training of the marginal model: $\epsilon= 10^{-2}$, and gradient clipping to norm/value with values 10/0.1 is applied, as for the marginal model.

\subsection{Results}

The quality of the trained model can be assessed using various measures, including the acceptance rate of samples used as proposals in independence Metropolis
or the Effective Sample Size per configuration (ESS), defined from $N$ samples as~\cite{Albergo:2021vyo} 
${\text{ESS } =\frac{1}{N} ( \sum_{i=1}^N w_i)^2\big/\sum_{i=1}^N  w_i^2 }$, with 
$w_i=p(\varphi_i)/q(\varphi_i)$.
ESS increases with model quality, with $\text{ESS} \in [1/N\, ,\, 1]$. With training as detailed above, the model achieves an ESS of approximately $0.1$ with 512 pseudofermion draws. 
For comparison, reweighting directly from the base distribution to the target gives an ESS of $\lesssim 10^{-4}$, estimated using 65k Haar-uniform configurations and exact evaluation of the determinant in the action.

\begin{figure*}[!t]
     \centering
     \subfloat[Plaquette\label{fig:plaq}]{%
     \includegraphics[width=0.49\textwidth]{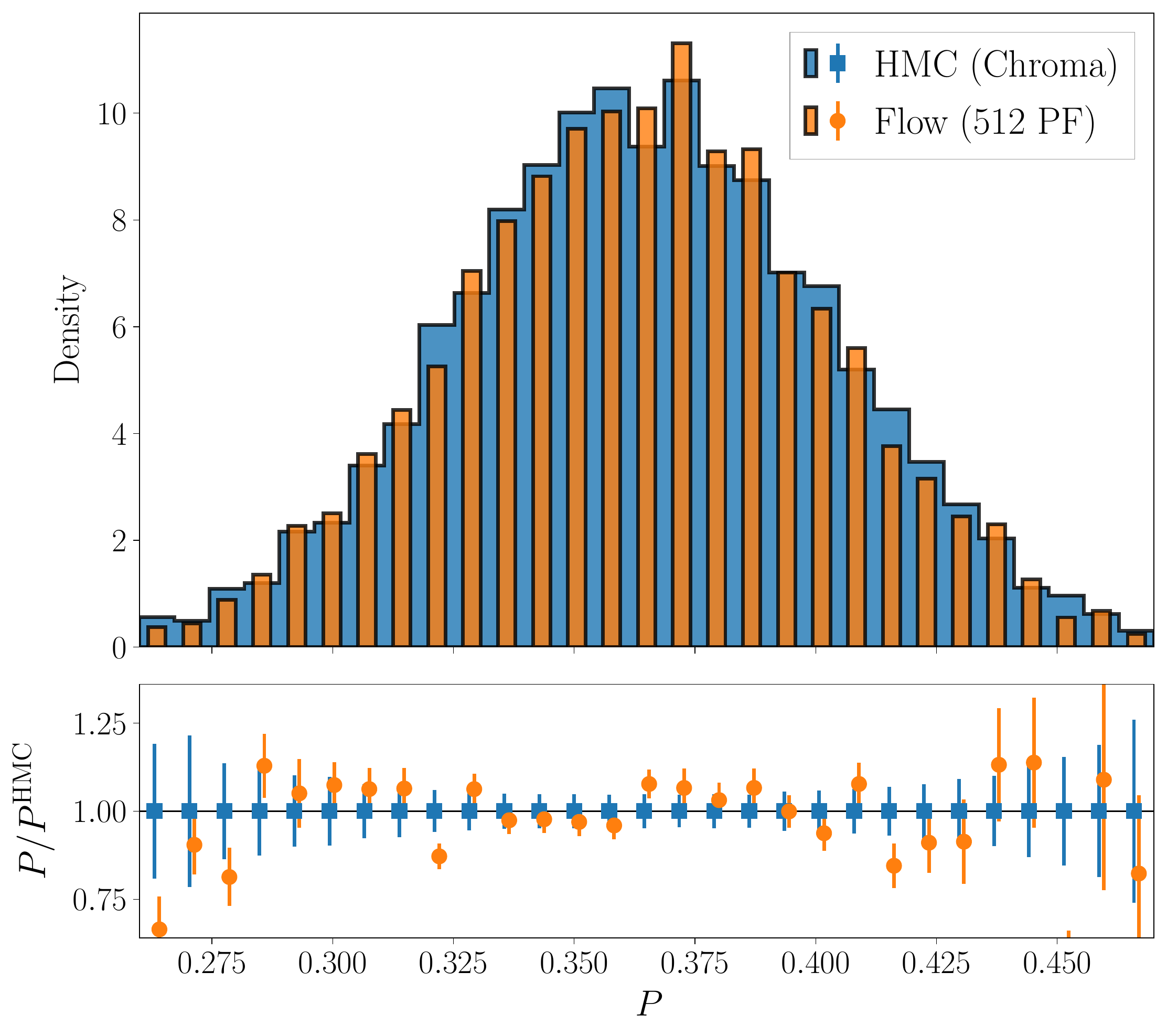}
    }
    \hfill
    \subfloat[Polyakov loop\label{fig:PL}]{%
     \includegraphics[width=0.49\textwidth]{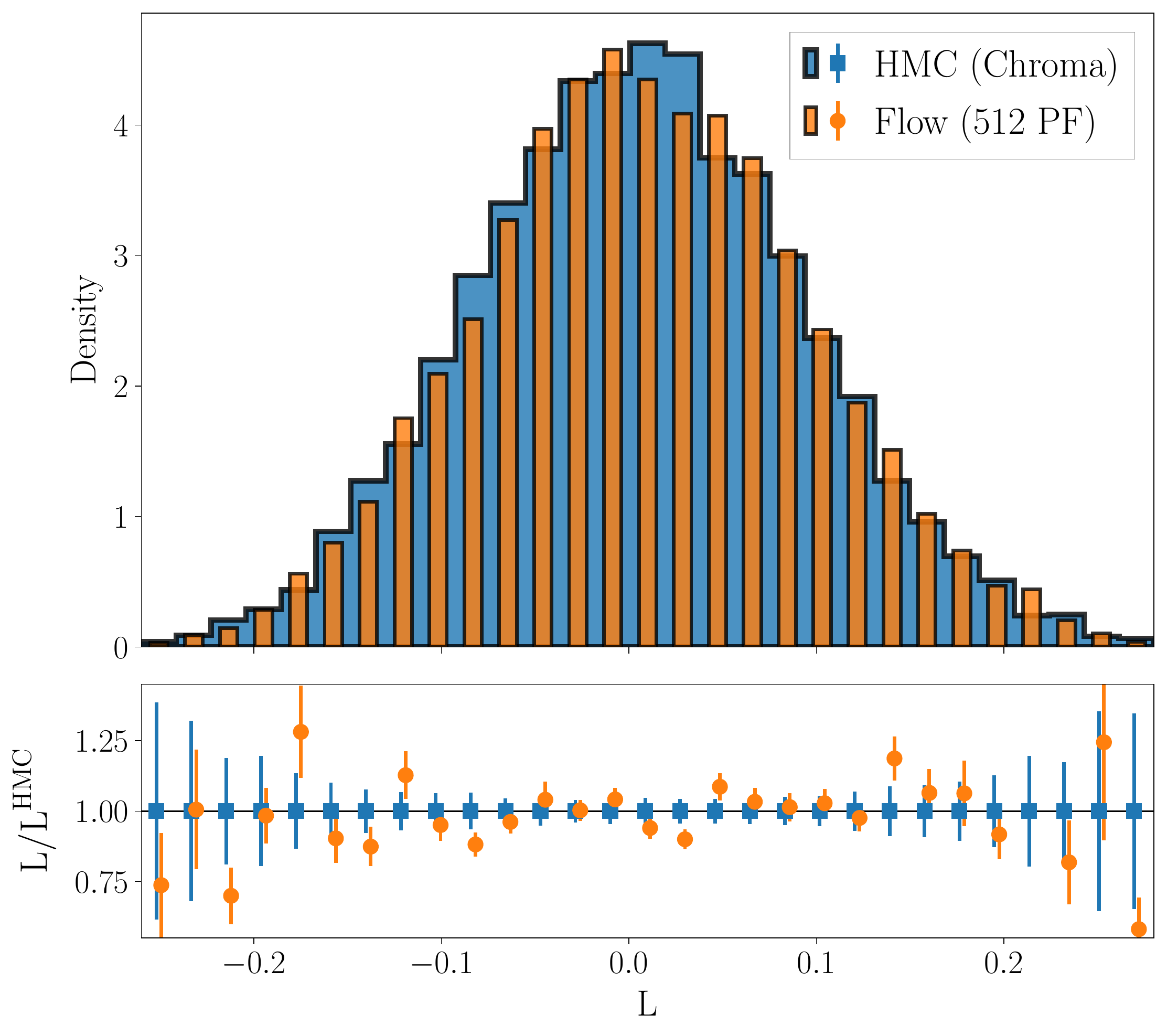}
    }\hfill \\ \centering \hfill
    \subfloat[Pion correlation function at $x_0=1$\label{fig:corr}]{%
     \includegraphics[width=0.49\textwidth]{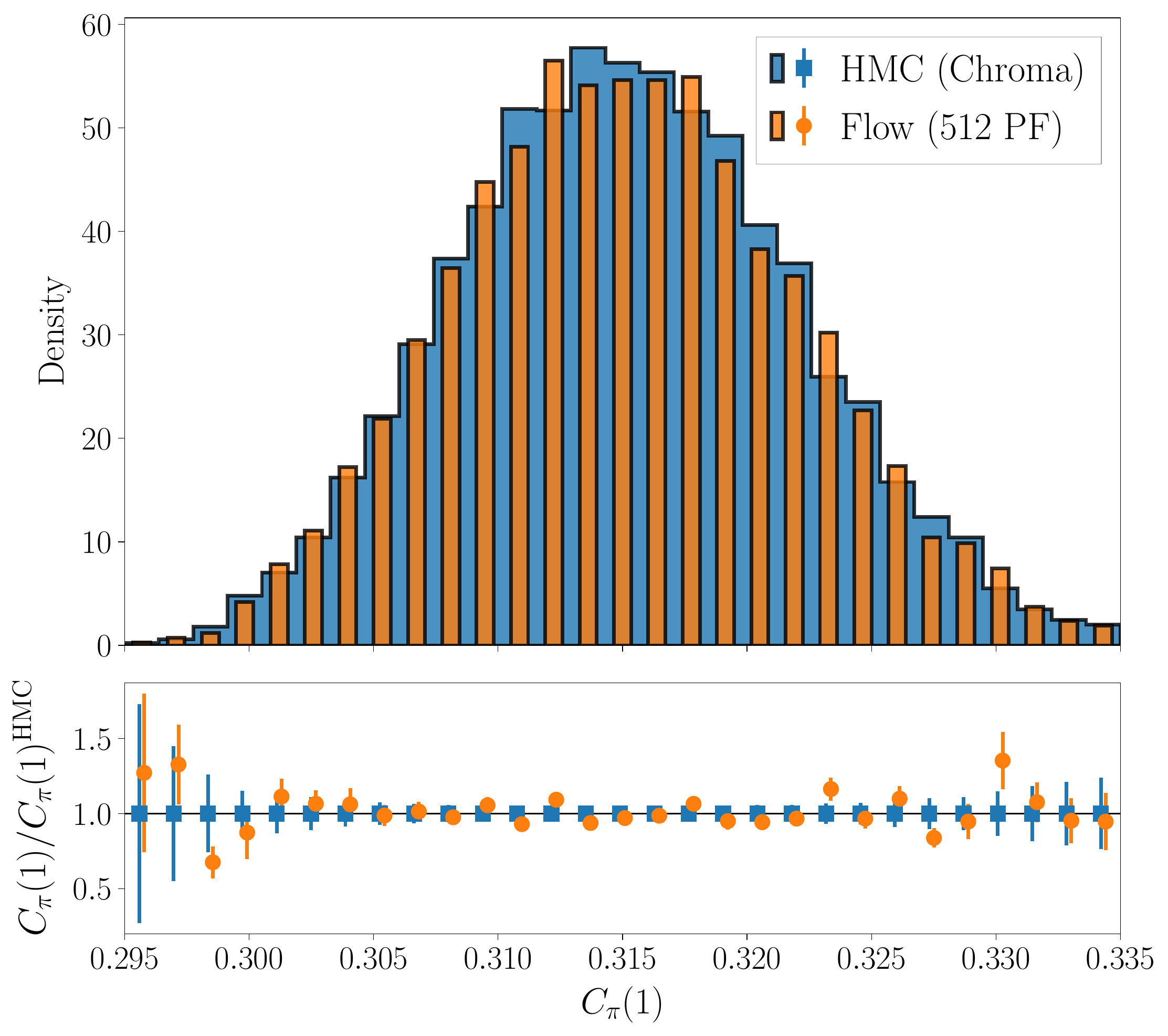}
    }
    \hfill
    \subfloat[Topological charge at $t/a^2=4$\label{fig:Q}]{%
     \includegraphics[width=0.49\textwidth]{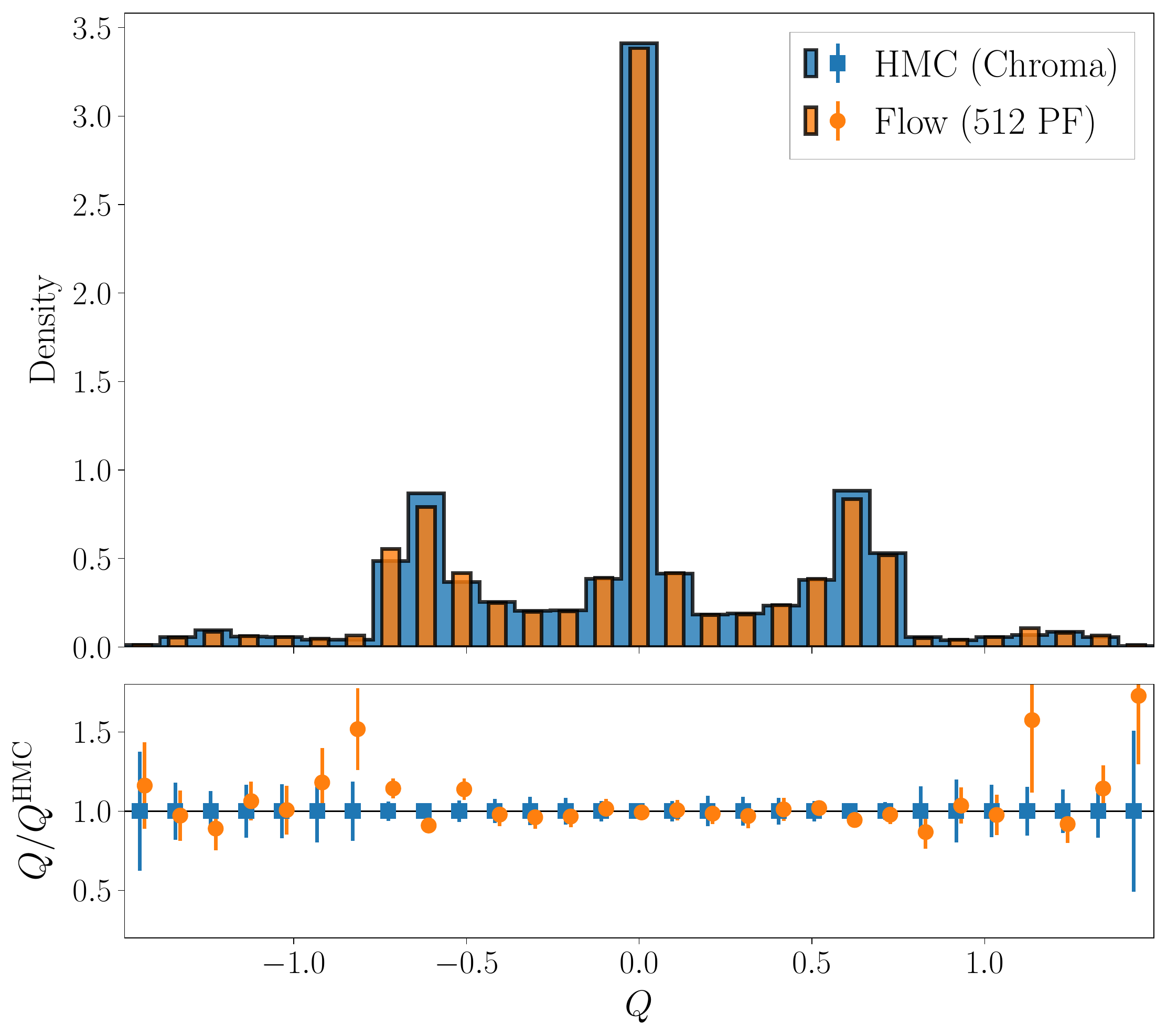}
    }
    \caption{Density histograms comparing various observables computed on ensembles of configurations sampled from the flow model, and generated via HMC, with equivalent statistics as described in the text. The lower panel of each subfigure shows the ratio of the counts in each bin of the histogram divided by the central value of the counts obtained in the same bin using the HMC ensemble; the uncertainties are estimated from 1000 bootstrap ensembles.}
    \label{fig:obs}
\end{figure*}

\begin{figure}[!t]
    \centering
    \includegraphics[width=0.6\linewidth]{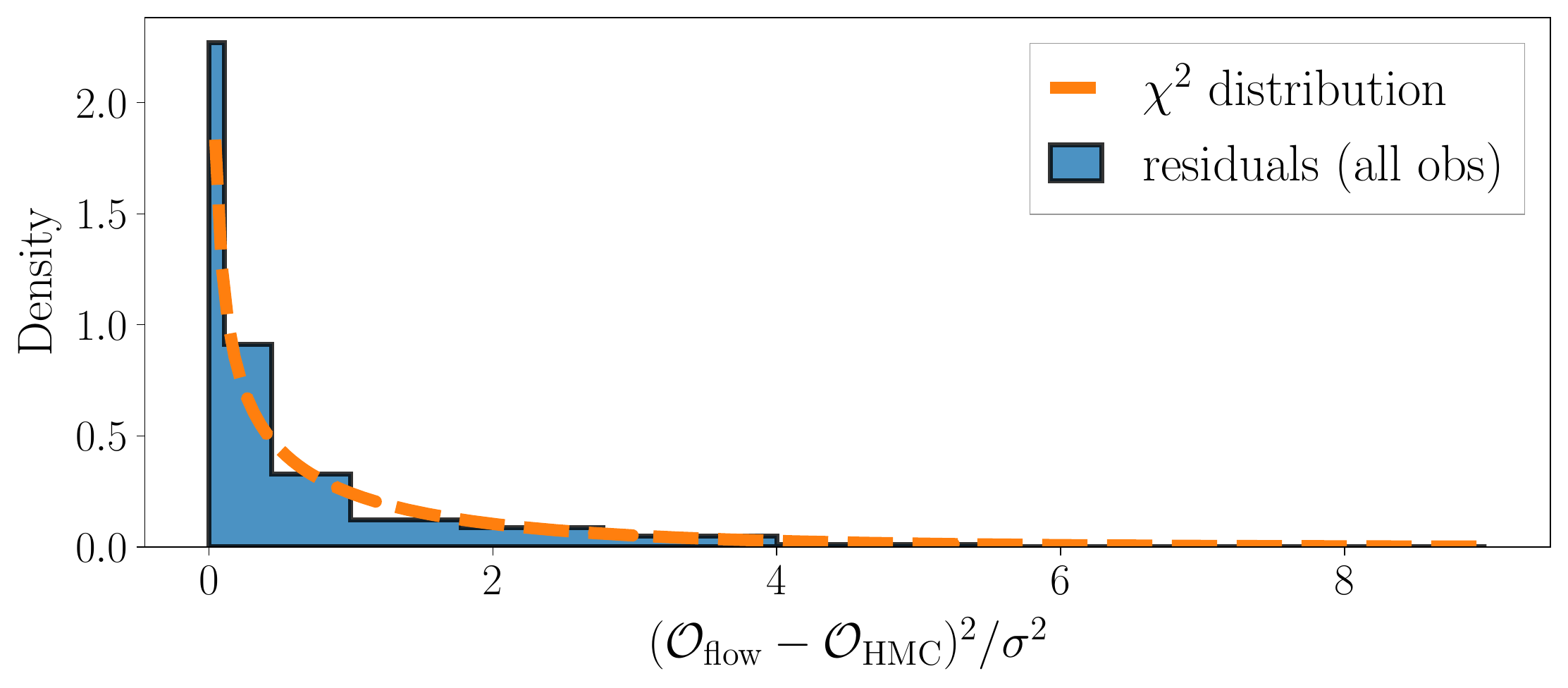}
    \caption{Distribution of the residuals of all the bins in the four subfigures in Fig.~\ref{fig:obs}. Uncertainties of the results obtained with the flow and HMC ensembles are added in quadrature. The dashed line illustrates a $\chi^2$ distribution with 1 degree of freedom. Binning is chosen to be uniform in the square root of the $x$-axis.}
    \label{fig:chi2}
\end{figure}

Comparing observables computed on an ensemble sampled from the trained flow model with those computed on an ensemble prepared using Hybrid Monte Carlo (HMC) at the same parameters shows statistical consistency between the algorithms, as is expected. 
Precisely, an ensemble of 65k configurations is sampled from the trained flow model; given the ESS of approximately 0.1, this corresponds to approximately 6.5k independent samples. Measurements on each configuration are reweighted when computing observables. For comparison, 65k configurations are generated for the same parameters using HMC as implemented in the {\tt chroma} library~\cite{Edwards:2004sx}. Trajectories of length $\tau = 0.5$ are integrated with five leapfrog steps each, yielding an acceptance rate of $\sim 95\%$. Configurations are generated in 8 independent streams; the initial 500 trajectories in each are discarded for equilibration. The ensemble is thinned by a factor of 10 to remove autocorrelations, for a total of 6.5k independent samples. 
Figure~\ref{fig:obs} collates a comparison of several observables computed on both ensembles:
\begin{itemize}
    \item Plaquette:
    \begin{equation}
        P = \frac{1}{N_c} \frac{1}{L^4} \sum_x \sum_{\mu < \nu }\text{Re } \text{tr }P_{\mu\nu}(x),
    \end{equation}
    where $N_c=3$ is the number of colors, $L=4$ is the extent of the lattice geometry, and $P_{\mu\nu}$ denotes the $1\times 1$ Wilson loop which extends in the $\mu$ and $\nu$ directions; 
    \item Polyakov loop:
    \begin{equation}
        L = \frac{1}{L^3} \sum_{\vec x}  \text{tr }  \prod_{x_0} U_0(x_0, \vec x),
    \end{equation}
    where $U_0$ is the gauge link in the time direction;
    \item Pion correlation function:
    \begin{equation}
        C_\pi(x_0) = - \sum_{\vec x}\langle [\bar u \gamma_5 d](x_0, \vec x) [\bar d \gamma_5 u](0,\vec{0})  \rangle,
    \end{equation}
    measured using point sources;
    \item Topological charge: 
    \begin{equation}
        Q = \frac{1}{16\pi^2} \sum_x \epsilon_{ \mu \nu \rho \sigma } F_{\mu\nu} (x) F_{\rho \sigma} (x),
    \end{equation}
    where $\epsilon_{ \mu \nu \rho \sigma }$ is the Levi-Civita tensor and $F_{\mu\nu}$ is defined through the clover definition. The topological charge is computed on ensembles subject to Wilson flow~\cite{Luscher:2009eq} to flow time $t/a^2=4$ in order to reveal the structure of the distribution.
\end{itemize}
Clearly, statistically consistent results are obtained with flow-based sampling as compared with HMC. The small discrepancies in some bins are consistent with statistical fluctuations, as illustrated in Fig.~\ref{fig:chi2}, which displays the distribution over all observables and all bins of residuals of the flow results relative to the HMC benchmark, as compared with the $\chi^2$ distribution.

\section{Outlook}

Section~\ref{sec:numericaldemo} demonstrates the basic technology for the application of flow-based sampling algorithms to lattice QCD. The key question, naturally, is whether any such approach will provide a computational advantage over HMC at the scale of state-of-the-art studies. As discussed at length in Ref.~\cite{scaling}, this is a challenging question that can only be answered through experimental investigation.   

The key difficulty in assessing the viability of flow-based sampling for lattice QCD is that the effectiveness of the approach depends on a very large number of choices, few of which have been explored to date. In particular, different flow architectures and training approaches yield not only different model quality and hence algorithm performance, but different scaling of that quality across parameter space. Moreover, there is a complicated trade-off between investment in training, and hence model quality, and the resulting acceleration of sampling; the optimal choice of stopping condition for optimization depends on how many configurations, at how many different sets of parameters, are ultimately desired, and may involve artificially truncating training of the model. To give just one example of the complexity of this question; incorporating the latest advances in flow model architecture design, flow-based sampling of scalar field theory configurations can be achieved that is at least an order of magnitude more efficient than the first implementation of the approach in Ref.~\cite{Albergo:2019eim}. For QCD in particular, while a first demonstration has now been achieved, the exploration of the space of suitable architectures has barely begun, and significant advances can be expected through further development.

Complicating the picture further, it is very likely that the first practical, at-scale applications of flow transformations in lattice QCD will not be in the context of direct sampling from the model, but via one of the many possible hybrid algorithms that can be developed to exploit the same or similar flow architectures, as noted in the introduction. Naturally, each such approach presents its own set of advantages and challenges with regard to scaling, which must be assessed independently.

Despite the unknowns, what is nevertheless evident is that the practical application of flow-based sampling in lattice QCD will require the feat of engineering a custom machine learning architecture to a similar scale as those routinely used in industrial applications. Moreover, the deployment of that model will likely require tight coupling with traditional lattice QCD codes on high-performance computing resources. Despite the clear challenge this presents, it is also clear that much stands to be gained; in particular, the approach could enable efficient ensemble generation at fine lattice spacings where HMC suffers from effects such as topological freezing and critical slowing-down, and offers clear advantages in studies of thermodynamic observables by providing direct access to the partition function. As a single trained model may be used to generate ensembles efficiently across parameter space, one can imagine using the same core model components to generate the diverse ensembles required for the control over the continuum limit in precision flavour physics applications, elongated lattice geometries or anisotropic field configurations for thermodynamics or spectroscopy, and high statistics to beat down signal-to-noise challenges in nuclear physics applications. If they can be achieved, optimized flow models for QCD are thus likely to become a valuable community resource, shared and re-used extensively for applications across lattice QCD.

\section*{Acknowledgements}
RA, DCH, GK, FRL, and PES are supported in part by the U.S.\ Department of Energy, Office of Science, Office of Nuclear Physics, under grant Contract Number DE-SC0011090. DCH, FRL and PES thank the Institute for Nuclear Theory at the University of Washington for its kind hospitality and stimulating research environment. This research was supported in part by the INT's U.S. Department of Energy grant No. DE-FG02-00ER41132. PES is additionally supported by the National Science Foundation under EAGER grant 2035015, by the U.S.\ DOE Early Career Award DE-SC0021006, by a NEC research award, and by the Carl G and Shirley Sontheimer Research Fund. GK is additionally supported by the Schweizerischer Nationalfonds. KC and MSA are supported by the National Science Foundation under the award PHY-2141336. MSA thanks the Flatiron Institute for their hospitality. DB is supported by the Argonne Leadership Computing Facility, which is a U.S. Department of Energy Office of Science User Facility operated under contract DE-AC02-06CH11357. This work is funded by the Deutsche Forschungsgemeinschaft (DFG, German Research Foundation) under Germany's Excellence Strategy EXC 2181/1 - 390900948 (the Heidelberg STRUCTURES Excellence Cluster), the Collaborative Research Centre SFB 1225 (ISOQUANT), and the U.S.\ National Science Foundation under Cooperative Agreement PHY-2019786 (The NSF AI Institute for Artificial Intelligence and Fundamental Interactions, \url{http://iaifi.org/}). This work is associated with an ALCF Aurora Early Science Program project, and used resources of the Argonne Leadership Computing Facility, which is a DOE Office of Science User Facility supported under Contract DE-AC02-06CH11357. The authors acknowledge the MIT SuperCloud and Lincoln Laboratory Supercomputing Center~\cite{reuther2018interactive} for providing HPC resources that have contributed to the research results reported within this paper. Numerical experiments and data analysis used PyTorch~\cite{NEURIPS2019_9015}, JAX~\cite{jax2018github}, Haiku~\cite{haiku2020github}, Horovod~\cite{sergeev2018horovod}, NumPy~\cite{harris2020array}, SciPy~\cite{2020SciPy-NMeth}, CHROMA~\cite{Edwards:2004sx}, and QUDA~\cite{Clark:2009wm}. Figures were produced using matplotlib~\cite{Hunter:2007}.

\bibliographystyle{utphys}
\bibliography{main}

\providecommand{\href}[2]{#2}\begingroup\raggedright\begin{thebibliography}{100}

\bibitem{Carrasquilla:2020mas}
J.~Carrasquilla \href{http://dx.doi.org/10.1080/23746149.2020.1797528}{{\em
  Adv. Phys. X} {\bfseries 5} no.~1, (2020) 1797528},
  \href{http://arxiv.org/abs/2003.11040}{{\ttfamily arXiv:2003.11040
  [physics.comp-ph]}}.

\bibitem{Boehnlein:2021eym}
A.~Boehnlein {\em et~al.} \href{http://arxiv.org/abs/2112.02309}{{\ttfamily
  arXiv:2112.02309 [nucl-th]}}.

\bibitem{Dawid:2022fga}
A.~Dawid {\em et~al.}
\newblock 4, 2022.
\newblock \href{http://arxiv.org/abs/2204.04198}{{\ttfamily arXiv:2204.04198
  [quant-ph]}}.

\bibitem{Boyda:2022nmh}
D.~Boyda {\em et~al.} in {\em {2022 Snowmass Summer Study}}.
\newblock 2, 2022.
\newblock \href{http://arxiv.org/abs/2202.05838}{{\ttfamily arXiv:2202.05838
  [hep-lat]}}.

\bibitem{Nicoli:2020njz}
K.~A. Nicoli, C.~J. Anders, L.~Funcke, T.~Hartung, K.~Jansen, P.~Kessel,
  S.~Nakajima, and P.~Stornati
  \href{http://dx.doi.org/10.1103/PhysRevLett.126.032001}{{\em Phys. Rev.
  Lett.} {\bfseries 126} no.~3, (2021) 032001},
  \href{http://arxiv.org/abs/2007.07115}{{\ttfamily arXiv:2007.07115
  [hep-lat]}}.

\bibitem{medvidovic2021generative}
M.~Medvidovic, J.~Carrasquilla, L.~E. Hayward, and B.~Kulchytskyy
  \href{http://arxiv.org/abs/2012.01442}{{\ttfamily arXiv:2012.01442
  [cond-mat.dis-nn]}}.

\bibitem{Foreman:2021ixr}
S.~Foreman, X.-Y. Jin, and J.~C. Osborn in {\em {9th International Conference
  on Learning Representations}}.
\newblock 5, 2021.
\newblock \href{http://arxiv.org/abs/2105.03418}{{\ttfamily arXiv:2105.03418
  [hep-lat]}}.

\bibitem{Foreman:2021ljl}
S.~Foreman, T.~Izubuchi, L.~Jin, X.-Y. Jin, J.~C. Osborn, and A.~Tomiya in {\em
  {38th International Symposium on Lattice Field Theory}}.
\newblock Dec, 2021.
\newblock \href{http://arxiv.org/abs/2112.01586}{{\ttfamily arXiv:2112.01586
  [cs.LG]}}.

\bibitem{Finkenrath:2022ogg}
J.~Finkenrath \href{http://arxiv.org/abs/2201.02216}{{\ttfamily
  arXiv:2201.02216 [hep-lat]}}.

\bibitem{Wang2017}
L.~Wang \href{http://dx.doi.org/10.1103/PhysRevE.96.051301}{{\em Phys. Rev. E}
  {\bfseries 96} (Nov, 2017) 051301}.
  \url{https://link.aps.org/doi/10.1103/PhysRevE.96.051301}.

\bibitem{Huang:2017}
L.~Huang and L.~Wang \href{http://dx.doi.org/10.1103/physrevb.95.035105}{{\em
  Physical Review B} {\bfseries 95} no.~3, (Jan, 2017) --}.
  \url{http://dx.doi.org/10.1103/PhysRevB.95.035105}.

\bibitem{song2017nice}
J.~Song, S.~Zhao, and S.~Ermon in {\em Advances in Neural Information
  Processing Systems}, pp.~5140--5150.
\newblock 2017.

\bibitem{Tanaka:2017niz}
A.~Tanaka and A.~Tomiya \href{http://arxiv.org/abs/1712.03893}{{\ttfamily
  arXiv:1712.03893 [hep-lat]}}.

\bibitem{levy2018generalizing}
D.~Levy, M.~D. Hoffman, and J.~Sohl-Dickstein
  \href{http://arxiv.org/abs/1711.09268}{{\ttfamily arXiv:1711.09268
  [stat.ML]}}.

\bibitem{Pawlowski:2018qxs}
J.~M. Pawlowski and J.~M. Urban
  \href{http://dx.doi.org/10.1088/2632-2153/abae73}{{\em Mach. Learn. Sci.
  Tech.} {\bfseries 1} (2020) 045011},
  \href{http://arxiv.org/abs/1811.03533}{{\ttfamily arXiv:1811.03533
  [hep-lat]}}.

\bibitem{Cossu:2018pxj}
G.~Cossu, L.~Del~Debbio, T.~Giani, A.~Khamseh, and M.~Wilson
  \href{http://dx.doi.org/10.1103/PhysRevB.100.064304}{{\em Phys. Rev. B}
  {\bfseries 100} no.~6, (2019) 064304},
  \href{http://arxiv.org/abs/1810.11503}{{\ttfamily arXiv:1810.11503
  [physics.comp-ph]}}.

\bibitem{Wu:2019}
D.~Wu, L.~Wang, and P.~Zhang
  \href{http://dx.doi.org/10.1103/PhysRevLett.122.080602}{{\em Phys. Rev.
  Lett.} {\bfseries 122} (Feb, 2019) 080602}.
  \url{https://link.aps.org/doi/10.1103/PhysRevLett.122.080602}.

\bibitem{Bachtis:2020dmf}
D.~Bachtis, G.~Aarts, and B.~Lucini
  \href{http://dx.doi.org/10.1103/PhysRevE.102.033303}{{\em Phys. Rev. E}
  {\bfseries 102} no.~3, (2020) 033303},
  \href{http://arxiv.org/abs/2004.14341}{{\ttfamily arXiv:2004.14341
  [cond-mat.stat-mech]}}.

\bibitem{Nagai:2020jar}
Y.~Nagai, A.~Tanaka, and A.~Tomiya
  \href{http://arxiv.org/abs/2010.11900}{{\ttfamily arXiv:2010.11900
  [hep-lat]}}.

\bibitem{Tomiya:2021ywc}
A.~Tomiya and Y.~Nagai \href{http://arxiv.org/abs/2103.11965}{{\ttfamily
  arXiv:2103.11965 [hep-lat]}}.

\bibitem{Bachtis:2021eww}
D.~Bachtis, G.~Aarts, F.~Di~Renzo, and B.~Lucini
  \href{http://dx.doi.org/10.1103/PhysRevLett.128.081603}{{\em Phys. Rev.
  Lett.} {\bfseries 128} no.~8, (2022) 081603},
  \href{http://arxiv.org/abs/2107.00466}{{\ttfamily arXiv:2107.00466
  [hep-lat]}}.

\bibitem{Wu:2021tfb}
D.~Wu, R.~Rossi, and G.~Carleo
  \href{http://dx.doi.org/10.1103/PhysRevResearch.3.L042024}{{\em Phys. Rev.
  Res.} {\bfseries 3} no.~4, (2021) L042024},
  \href{http://arxiv.org/abs/2105.05650}{{\ttfamily arXiv:2105.05650
  [cond-mat.stat-mech]}}.

\bibitem{Gerdes:2022eve}
M.~Gerdes, P.~de~Haan, C.~Rainone, R.~Bondesan, and M.~C.~N. Cheng
  \href{http://arxiv.org/abs/2207.00283}{{\ttfamily arXiv:2207.00283
  [hep-lat]}}.

\bibitem{Alexandru:2017czx}
A.~Alexandru, P.~F. Bedaque, H.~Lamm, and S.~Lawrence
  \href{http://dx.doi.org/10.1103/PhysRevD.96.094505}{{\em Phys. Rev. D}
  {\bfseries 96} no.~9, (2017) 094505},
  \href{http://arxiv.org/abs/1709.01971}{{\ttfamily arXiv:1709.01971
  [hep-lat]}}.

\bibitem{Lawrence:2021izu}
S.~Lawrence and Y.~Yamauchi
  \href{http://dx.doi.org/10.1103/PhysRevD.103.114509}{{\em Phys. Rev. D}
  {\bfseries 103} no.~11, (2021) 114509},
  \href{http://arxiv.org/abs/2101.05755}{{\ttfamily arXiv:2101.05755
  [hep-lat]}}.

\bibitem{Wynen:2020uzx}
J.-L. Wynen, E.~Berkowitz, S.~Krieg, T.~Luu, and J.~Ostmeyer
  \href{http://dx.doi.org/10.1103/PhysRevB.103.125153}{{\em Phys. Rev. B}
  {\bfseries 103} no.~12, (2021) 125153},
  \href{http://arxiv.org/abs/2006.11221}{{\ttfamily arXiv:2006.11221
  [cond-mat.str-el]}}.

\bibitem{LiWang2018NNRG}
S.-H. Li and L.~Wang
  \href{http://dx.doi.org/10.1103/PhysRevLett.121.260601}{{\em Phys. Rev.
  Lett.} {\bfseries 121} (Dec, 2018) 260601}.
  \url{https://link.aps.org/doi/10.1103/PhysRevLett.121.260601}.

\bibitem{Albergo:2019eim}
M.~S. Albergo, G.~Kanwar, and P.~E. Shanahan
  \href{http://dx.doi.org/10.1103/PhysRevD.100.034515}{{\em Phys. Rev. D}
  {\bfseries 100} no.~3, (2019) 034515},
  \href{http://arxiv.org/abs/1904.12072}{{\ttfamily arXiv:1904.12072
  [hep-lat]}}.

\bibitem{Nicoli:2020evf}
K.~A. Nicoli, S.~Nakajima, N.~Strodthoff, W.~Samek, K.-R. M\"uller, and
  P.~Kessel \href{http://dx.doi.org/10.1103/PhysRevE.101.023304}{{\em Phys.
  Rev. E} {\bfseries 101} no.~2, (2020) 023304},
  \href{http://arxiv.org/abs/1910.13496}{{\ttfamily arXiv:1910.13496
  [cond-mat.stat-mech]}}.

\bibitem{Kanwar:2020xzo}
G.~Kanwar, M.~S. Albergo, D.~Boyda, K.~Cranmer, D.~C. Hackett, S.~Racani\`ere,
  D.~J. Rezende, and P.~E. Shanahan
  \href{http://dx.doi.org/10.1103/PhysRevLett.125.121601}{{\em Phys. Rev.
  Lett.} {\bfseries 125} no.~12, (2020) 121601},
  \href{http://arxiv.org/abs/2003.06413}{{\ttfamily arXiv:2003.06413
  [hep-lat]}}.

\bibitem{Boyda:2020hsi}
D.~Boyda, G.~Kanwar, S.~Racani\`ere, D.~J. Rezende, M.~S. Albergo, K.~Cranmer,
  D.~C. Hackett, and P.~E. Shanahan
  \href{http://dx.doi.org/10.1103/PhysRevD.103.074504}{{\em Phys. Rev. D}
  {\bfseries 103} no.~7, (2021) 074504},
  \href{http://arxiv.org/abs/2008.05456}{{\ttfamily arXiv:2008.05456
  [hep-lat]}}.

\bibitem{Albergo:2021bna}
M.~S. Albergo, G.~Kanwar, S.~Racani\`ere, D.~J. Rezende, J.~M. Urban, D.~Boyda,
  K.~Cranmer, D.~C. Hackett, and P.~E. Shanahan
  \href{http://dx.doi.org/10.1103/PhysRevD.104.114507}{{\em Phys. Rev. D}
  {\bfseries 104} no.~11, (2021) 114507},
  \href{http://arxiv.org/abs/2106.05934}{{\ttfamily arXiv:2106.05934
  [hep-lat]}}.

\bibitem{Hackett:2021idh}
D.~C. Hackett, C.-C. Hsieh, M.~S. Albergo, D.~Boyda, J.-W. Chen, K.-F. Chen,
  K.~Cranmer, G.~Kanwar, and P.~E. Shanahan
  \href{http://arxiv.org/abs/2107.00734}{{\ttfamily arXiv:2107.00734
  [hep-lat]}}.

\bibitem{Gabrie:2021tlu}
M.~Gabri\'e, G.~M. Rotskoff, and E.~Vanden-Eijnden
  \href{http://dx.doi.org/10.1073/pnas.2109420119}{{\em Proc. Nat. Acad. Sci.}
  {\bfseries 119} no.~10, (2022) e2109420119},
  \href{http://arxiv.org/abs/2105.12603}{{\ttfamily arXiv:2105.12603
  [physics.data-an]}}.

\bibitem{DelDebbio:2021qwf}
L.~Del~Debbio, J.~M. Rossney, and M.~Wilson
  \href{http://arxiv.org/abs/2105.12481}{{\ttfamily arXiv:2105.12481
  [hep-lat]}}.

\bibitem{Foreman:2021rhs}
S.~Foreman, X.-Y. Jin, and J.~C. Osborn
  \href{http://dx.doi.org/10.22323/1.396.0508}{{\em PoS} {\bfseries
  LATTICE2021} (2022) 508}, \href{http://arxiv.org/abs/2112.01582}{{\ttfamily
  arXiv:2112.01582 [hep-lat]}}.

\bibitem{Jin:2022bgq}
X.-Y. Jin \href{http://dx.doi.org/10.22323/1.396.0600}{{\em PoS} {\bfseries
  LATTICE2021} (2022) 600}, \href{http://arxiv.org/abs/2201.01862}{{\ttfamily
  arXiv:2201.01862 [hep-lat]}}.

\bibitem{deHaan:2021erb}
P.~de~Haan, C.~Rainone, M.~C.~N. Cheng, and R.~Bondesan
  \href{http://arxiv.org/abs/2110.02673}{{\ttfamily arXiv:2110.02673 [cs.LG]}}.

\bibitem{Singha:2022lpi}
A.~Singha, D.~Chakrabarti, and V.~Arora
  \href{http://arxiv.org/abs/2207.00980}{{\ttfamily arXiv:2207.00980
  [hep-lat]}}.

\bibitem{Matthews:2022sds}
A.~G. D.~G. Matthews, M.~Arbel, D.~J. Rezende, and A.~Doucet
  \href{http://arxiv.org/abs/2201.13117}{{\ttfamily arXiv:2201.13117
  [stat.ML]}}.

\bibitem{Caselle:2022acb}
M.~Caselle, E.~Cellini, A.~Nada, and M.~Panero
  \href{http://arxiv.org/abs/2201.08862}{{\ttfamily arXiv:2201.08862
  [hep-lat]}}.

\bibitem{Pawlowski:2022rdn}
J.~M. Pawlowski and J.~M. Urban
  \href{http://arxiv.org/abs/2203.01243}{{\ttfamily arXiv:2203.01243
  [hep-lat]}}.

\bibitem{Albergo:2022qfi}
M.~S. Albergo, D.~Boyda, K.~Cranmer, D.~C. Hackett, G.~Kanwar, S.~Racani\`ere,
  D.~J. Rezende, F.~Romero-L\'opez, P.~E. Shanahan, and J.~M. Urban
  \href{http://arxiv.org/abs/2202.11712}{{\ttfamily arXiv:2202.11712
  [hep-lat]}}.

\bibitem{Albergo:2021vyo}
M.~S. Albergo, D.~Boyda, D.~C. Hackett, G.~Kanwar, K.~Cranmer, S.~Racani\`ere,
  D.~J. Rezende, and P.~E. Shanahan
  \href{http://arxiv.org/abs/2101.08176}{{\ttfamily arXiv:2101.08176
  [hep-lat]}}.

\bibitem{Favoni:2020reg}
M.~Favoni, A.~Ipp, D.~I. M\"uller, and D.~Schuh
  \href{http://dx.doi.org/10.1103/PhysRevLett.128.032003}{{\em Phys. Rev.
  Lett.} {\bfseries 128} no.~3, (2022) 032003},
  \href{http://arxiv.org/abs/2012.12901}{{\ttfamily arXiv:2012.12901
  [hep-lat]}}.

\bibitem{Abbott:2022zhs}
R.~Abbott {\em et~al.} \href{http://arxiv.org/abs/2207.08945}{{\ttfamily
  arXiv:2207.08945 [hep-lat]}}.

\bibitem{Boyda:2020nfh}
D.~L. Boyda, M.~N. Chernodub, N.~V. Gerasimeniuk, V.~A. Goy, S.~D. Liubimov,
  and A.~V. Molochkov \href{http://dx.doi.org/10.1103/PhysRevD.103.014509}{{\em
  Phys. Rev. D} {\bfseries 103} no.~1, (2021) 014509},
  \href{http://arxiv.org/abs/2009.10971}{{\ttfamily arXiv:2009.10971
  [hep-lat]}}.

\bibitem{Bachtis:2020fly}
D.~Bachtis, G.~Aarts, and B.~Lucini
  \href{http://dx.doi.org/10.1103/PhysRevResearch.3.013134}{{\em Phys. Rev.
  Res.} {\bfseries 3} no.~1, (2021) 013134},
  \href{http://arxiv.org/abs/2010.00054}{{\ttfamily arXiv:2010.00054
  [hep-lat]}}.

\bibitem{Palermo:2021jrf}
A.~Palermo, L.~Anderlini, M.~P. Lombardo, A.~Y. Kotov, and A.~Trunin
  \href{http://dx.doi.org/10.22323/1.396.0030}{{\em PoS} {\bfseries
  LATTICE2021} (2022) 030}, \href{http://arxiv.org/abs/2111.05216}{{\ttfamily
  arXiv:2111.05216 [hep-lat]}}.

\bibitem{Tan:2021cgs}
D.~R. Tan, J.~H. Peng, Y.~H. Tseng, and F.~J. Jiang
  \href{http://dx.doi.org/10.1140/epjp/s13360-021-02121-4}{{\em Eur. Phys. J.
  Plus} {\bfseries 136} no.~11, (2021) 1116},
  \href{http://arxiv.org/abs/2103.10846}{{\ttfamily arXiv:2103.10846
  [cond-mat.dis-nn]}}.

\bibitem{Li:2017xaz}
C.-D. Li, D.-R. Tan, and F.-J. Jiang
  \href{http://dx.doi.org/10.1016/j.aop.2018.02.018}{{\em Annals Phys.}
  {\bfseries 391} (2018) 312--331},
  \href{http://arxiv.org/abs/1703.02369}{{\ttfamily arXiv:1703.02369
  [cond-mat.dis-nn]}}.

\bibitem{Wetzel:2017ooo}
S.~J. Wetzel and M.~Scherzer
  \href{http://dx.doi.org/10.1103/PhysRevB.96.184410}{{\em Phys. Rev. B}
  {\bfseries 96} no.~18, (2017) 184410},
  \href{http://arxiv.org/abs/1705.05582}{{\ttfamily arXiv:1705.05582
  [cond-mat.stat-mech]}}.

\bibitem{Alexandrou:2019hgt}
C.~Alexandrou, A.~Athenodorou, C.~Chrysostomou, and S.~Paul
  \href{http://dx.doi.org/10.1140/epjb/e2020-100506-5}{{\em Eur. Phys. J. B}
  {\bfseries 93} no.~12, (2020) 226},
  \href{http://arxiv.org/abs/1903.03506}{{\ttfamily arXiv:1903.03506
  [cond-mat.stat-mech]}}.

\bibitem{Blucher:2020mjt}
S.~Bl\"ucher, L.~Kades, J.~M. Pawlowski, N.~Strodthoff, and J.~M. Urban
  \href{http://dx.doi.org/10.1103/PhysRevD.101.094507}{{\em Phys. Rev. D}
  {\bfseries 101} no.~9, (2020) 094507},
  \href{http://arxiv.org/abs/2003.01504}{{\ttfamily arXiv:2003.01504
  [hep-lat]}}.

\bibitem{Yau:2020emg}
H.~M. Yau and N.~Su
  \href{http://dx.doi.org/10.21468/SciPostPhysCore.5.2.032}{{\em SciPost Phys.
  Core} {\bfseries 5} (2022) 032},
  \href{http://arxiv.org/abs/2006.15021}{{\ttfamily arXiv:2006.15021
  [cond-mat.dis-nn]}}.

\bibitem{Zhou:2018ill}
K.~Zhou, G.~Endr\H{o}di, L.-G. Pang, and H.~St\"ocker
  \href{http://dx.doi.org/10.1103/PhysRevD.100.011501}{{\em Phys. Rev. D}
  {\bfseries 100} no.~1, (2019) 011501},
  \href{http://arxiv.org/abs/1810.12879}{{\ttfamily arXiv:1810.12879
  [hep-lat]}}.

\bibitem{Detmold:2021ulb}
W.~Detmold, G.~Kanwar, H.~Lamm, M.~L. Wagman, and N.~C. Warrington
  \href{http://dx.doi.org/10.1103/PhysRevD.103.094517}{{\em Phys. Rev. D}
  {\bfseries 103} no.~9, (2021) 094517},
  \href{http://arxiv.org/abs/2101.12668}{{\ttfamily arXiv:2101.12668
  [hep-lat]}}.

\bibitem{hu2017discovering}
W.~Hu, R.~R. Singh, and R.~T. Scalettar
  \href{http://dx.doi.org/10.1103/PhysRevE.95.062122}{{\em Physical Review E}
  {\bfseries 95} no.~6, (2017) 062122},
  \href{http://arxiv.org/abs/1704.00080}{{\ttfamily arXiv:1704.00080
  [cond-mat.stat-mech]}}.

\bibitem{wetzel2020discovering}
S.~J. Wetzel, R.~G. Melko, J.~Scott, M.~Panju, and V.~Ganesh
  \href{http://dx.doi.org/10.1103/physrevresearch.2.033499}{{\em Phys. Rev.
  Res.} {\bfseries 2} no.~3, (2020) 033499},
  \href{http://arxiv.org/abs/2003.04299}{{\ttfamily arXiv:2003.04299
  [physics.comp-ph]}}.

\bibitem{Yoon:2018krb}
B.~Yoon, T.~Bhattacharya, and R.~Gupta
  \href{http://dx.doi.org/10.1103/PhysRevD.100.014504}{{\em Phys. Rev. D}
  {\bfseries 100} no.~1, (2019) 014504},
  \href{http://arxiv.org/abs/1807.05971}{{\ttfamily arXiv:1807.05971
  [hep-lat]}}.

\bibitem{Zhang:2019qiq}
R.~Zhang, Z.~Fan, R.~Li, H.-W. Lin, and B.~Yoon
  \href{http://dx.doi.org/10.1103/PhysRevD.101.034516}{{\em Phys. Rev. D}
  {\bfseries 101} no.~3, (2020) 034516},
  \href{http://arxiv.org/abs/1909.10990}{{\ttfamily arXiv:1909.10990
  [hep-lat]}}.

\bibitem{Shanahan:2018vcv}
P.~E. Shanahan, D.~Trewartha, and W.~Detmold
  \href{http://dx.doi.org/10.1103/PhysRevD.97.094506}{{\em Phys. Rev. D}
  {\bfseries 97} no.~9, (2018) 094506},
  \href{http://arxiv.org/abs/1801.05784}{{\ttfamily arXiv:1801.05784
  [hep-lat]}}.

\bibitem{Hudspith:2021iqu}
R.~J. Hudspith and D.~Mohler \href{http://arxiv.org/abs/2112.01997}{{\ttfamily
  arXiv:2112.01997 [hep-lat]}}.

\bibitem{Kades:2019wtd}
L.~Kades, J.~M. Pawlowski, A.~Rothkopf, M.~Scherzer, J.~M. Urban, S.~J. Wetzel,
  N.~Wink, and F.~P.~G. Ziegler
  \href{http://dx.doi.org/10.1103/PhysRevD.102.096001}{{\em Phys. Rev. D}
  {\bfseries 102} no.~9, (2020) 096001},
  \href{http://arxiv.org/abs/1905.04305}{{\ttfamily arXiv:1905.04305
  [physics.comp-ph]}}.

\bibitem{Offler:2019eij}
S.~Offler, G.~Aarts, C.~Allton, J.~Glesaaen, B.~J\"ager, S.~Kim, M.~P.
  Lombardo, S.~M. Ryan, and J.-I. Skullerud
  \href{http://dx.doi.org/10.22323/1.363.0076}{{\em PoS} {\bfseries
  LATTICE2019} (2019) 076}, \href{http://arxiv.org/abs/1912.12900}{{\ttfamily
  arXiv:1912.12900 [hep-lat]}}.

\bibitem{Horak:2021syv}
J.~Horak, J.~M. Pawlowski, J.~Rodr\'\i{}guez-Quintero, J.~Turnwald, J.~M.
  Urban, N.~Wink, and S.~Zafeiropoulos
  \href{http://dx.doi.org/10.1103/PhysRevD.105.036014}{{\em Phys. Rev. D}
  {\bfseries 105} no.~3, (2022) 036014},
  \href{http://arxiv.org/abs/2107.13464}{{\ttfamily arXiv:2107.13464
  [hep-ph]}}.

\bibitem{Chen:2021giw}
S.~Y. Chen, H.~T. Ding, F.~Y. Liu, G.~Papp, and C.~B. Yang
  \href{http://arxiv.org/abs/2110.13521}{{\ttfamily arXiv:2110.13521
  [hep-lat]}}.

\bibitem{Wang:2021cqw}
L.~Wang, S.~Shi, and K.~Zhou in {\em {35th Conference on Neural Information
  Processing Systems}}.
\newblock 12, 2021.
\newblock \href{http://arxiv.org/abs/2112.06206}{{\ttfamily arXiv:2112.06206
  [physics.comp-ph]}}.

\bibitem{Shi:2022yqw}
S.~Shi, L.~Wang, and K.~Zhou \href{http://arxiv.org/abs/2201.02564}{{\ttfamily
  arXiv:2201.02564 [hep-ph]}}.

\bibitem{DelDebbio:2020rgv}
L.~Del~Debbio, T.~Giani, J.~Karpie, K.~Orginos, A.~Radyushkin, and
  S.~Zafeiropoulos \href{http://dx.doi.org/10.1007/JHEP02(2021)138}{{\em JHEP}
  {\bfseries 02} (2021) 138}, \href{http://arxiv.org/abs/2010.03996}{{\ttfamily
  arXiv:2010.03996 [hep-ph]}}.

\bibitem{Karpie:2019eiq}
J.~Karpie, K.~Orginos, A.~Rothkopf, and S.~Zafeiropoulos
  \href{http://dx.doi.org/10.1007/JHEP04(2019)057}{{\em JHEP} {\bfseries 04}
  (2019) 057}, \href{http://arxiv.org/abs/1901.05408}{{\ttfamily
  arXiv:1901.05408 [hep-lat]}}.

\bibitem{Luscher:2009eq}
M.~Lüscher \href{http://dx.doi.org/10.1007/s00220-009-0953-7}{{\em Commun.
  Math. Phys.} {\bfseries 293} (2010) 899--919},
  \href{http://arxiv.org/abs/0907.5491}{{\ttfamily arXiv:0907.5491 [hep-lat]}}.

\bibitem{rezende2016variational}
D.~J. Rezende and S.~Mohamed \href{http://arxiv.org/abs/1505.05770}{{\ttfamily
  arXiv:1505.05770 [stat.ML]}}.

\bibitem{dinh2017density}
L.~Dinh, J.~Sohl-Dickstein, and S.~Bengio
  \href{http://arxiv.org/abs/1605.08803}{{\ttfamily arXiv:1605.08803 [cs.LG]}}.

\bibitem{JMLR:v22:19-1028}
G.~Papamakarios, E.~Nalisnick, D.~J. Rezende, S.~Mohamed, and
  B.~Lakshminarayanan {\em Journal of Machine Learning Research} {\bfseries 22}
  no.~57, (2021) 1--64.

\bibitem{Nicoli:2021inv}
K.~A. Nicoli, C.~J. Anders, L.~Funcke, T.~Hartung, K.~Jansen, P.~Kessel,
  S.~Nakajima, and P.~Stornati
  \href{http://dx.doi.org/10.22323/1.396.0338}{{\em PoS} {\bfseries
  LATTICE2021} (2022) 338}, \href{http://arxiv.org/abs/2111.11303}{{\ttfamily
  arXiv:2111.11303 [hep-lat]}}.

\bibitem{Kullback:1951}
S.~Kullback and R.~A. Leibler
  \href{http://dx.doi.org/10.1214/aoms/1177729694}{{\em The Annals of
  Mathematical Statistics} {\bfseries 22} no.~1, (1951) 79 -- 86}.

\bibitem{Metropolis:1953am}
N.~Metropolis, A.~W. Rosenbluth, M.~N. Rosenbluth, A.~H. Teller, and E.~Teller
  \href{http://dx.doi.org/10.1063/1.1699114}{{\em J. Chem. Phys.} {\bfseries
  21} (1953) 1087--1092}.

\bibitem{Hastings:1970aa}
W.~K. Hastings \href{http://dx.doi.org/10.1093/biomet/57.1.97}{{\em Biometrika}
  {\bfseries 57} (1970) 97--109}.

\bibitem{tierney1994markov}
L.~Tierney {\em the Annals of Statistics} (1994) 1701--1728.

\bibitem{Creutz:1979dw}
M.~Creutz \href{http://dx.doi.org/10.1103/PhysRevLett.43.553}{{\em Phys. Rev.
  Lett.} {\bfseries 43} (1979) 553--556}. [Erratum: Phys.Rev.Lett. 43, 890
  (1979)].

\bibitem{Cabibbo:1982zn}
N.~Cabibbo and E.~Marinari
  \href{http://dx.doi.org/10.1016/0370-2693(82)90696-7}{{\em Phys. Lett. B}
  {\bfseries 119} (1982) 387--390}.

\bibitem{Kennedy:1985nu}
A.~D. Kennedy and B.~J. Pendleton
  \href{http://dx.doi.org/10.1016/0370-2693(85)91632-6}{{\em Phys. Lett. B}
  {\bfseries 156} (1985) 393--399}.

\bibitem{Yamauchi:2021kpo}
Y.~Yamauchi and S.~Lawrence \href{http://dx.doi.org/10.22323/1.396.0621}{{\em
  PoS} {\bfseries LATTICE2021} (2022) 621},
  \href{http://arxiv.org/abs/2112.15035}{{\ttfamily arXiv:2112.15035
  [hep-lat]}}.

\bibitem{Lawrence:2022afv}
S.~Lawrence, H.~Oh, and Y.~Yamauchi
  \href{http://arxiv.org/abs/2205.12303}{{\ttfamily arXiv:2205.12303
  [hep-lat]}}.

\bibitem{higherdim}
R.~Abbott {\em et~al.} {\em Flow-based sampling for lattice gauge theories in
  higher dimensions (in preparation)} (2022) .

\bibitem{He:2015dtg}
K.~He, X.~Zhang, S.~Ren, and J.~Sun
  \href{http://arxiv.org/abs/1502.01852}{{\ttfamily arXiv:1502.01852 [cs.CV]}}.

\bibitem{kingma2017adam}
D.~P. Kingma and J.~Ba \href{http://arxiv.org/abs/1412.6980}{{\ttfamily
  arXiv:1412.6980 [cs.LG]}}.

\bibitem{zhang2020gradient}
J.~Zhang, T.~He, S.~Sra, and A.~Jadbabaie
  \href{http://arxiv.org/abs/1905.11881}{{\ttfamily arXiv:1905.11881
  [math.OC]}}.

\bibitem{Edwards:2004sx}
{SciDAC, LHPC, UKQCD} Collaboration, R.~G. Edwards and B.~Joo
  \href{http://dx.doi.org/10.1016/j.nuclphysbps.2004.11.254}{{\em Nucl. Phys. B
  Proc. Suppl.} {\bfseries 140} (2005) 832},
  \href{http://arxiv.org/abs/hep-lat/0409003}{{\ttfamily
  arXiv:hep-lat/0409003}}.

\bibitem{scaling}
R.~Abbott {\em et~al.} {\em Aspects of scaling and scalability for flow-based
  sampling of lattice QCD (in preparation)} (2022) .

\bibitem{reuther2018interactive}
A.~Reuther, J.~Kepner, C.~Byun, S.~Samsi, W.~Arcand, D.~Bestor, B.~Bergeron,
  V.~Gadepally, M.~Houle, M.~Hubbell, {\em et~al.}
  \href{http://dx.doi.org/10.1109/hpec.2018.8547629}{{\em 2018 IEEE High
  Performance extreme Computing Conference (HPEC)} (Sep, 2018) 1--6},
  \href{http://arxiv.org/abs/1807.07814}{{\ttfamily arXiv:1807.07814 [cs.DC]}}.

\bibitem{NEURIPS2019_9015}
A.~Paszke {\em et~al.} in {\em Advances in Neural Information Processing
  Systems 32}, H.~Wallach, H.~Larochelle, A.~Beygelzimer, F.~d\textquotesingle
  Alch\'{e}-Buc, E.~Fox, and R.~Garnett, eds., pp.~8024--8035.
\newblock Curran Associates, Inc., 2019.
\newblock
  \url{http://papers.neurips.cc/paper/9015-pytorch-an-imperative-style-high-performance\\-deep-learning-library.pdf}.

\bibitem{jax2018github}
J.~Bradbury, R.~Frostig, P.~Hawkins, M.~J. Johnson, C.~Leary, D.~Maclaurin,
  G.~Necula, A.~Paszke, J.~Vander{P}las, S.~Wanderman-{M}ilne, and Q.~Zhang,
  2018.
\newblock \url{http://github.com/google/jax}.

\bibitem{haiku2020github}
T.~Hennigan, T.~Cai, T.~Norman, and I.~Babuschkin, 2020.
\newblock \url{http://github.com/deepmind/dm-haiku}.

\bibitem{sergeev2018horovod}
A.~{Sergeev} and M.~{Del Balso}
  \href{http://arxiv.org/abs/1802.05799}{{\ttfamily arXiv:1802.05799 [cs.LG]}}.

\bibitem{harris2020array}
C.~R. Harris, K.~J. Millman, S.~J. Van Der~Walt, R.~Gommers, P.~Virtanen,
  D.~Cournapeau, E.~Wieser, J.~Taylor, S.~Berg, N.~J. Smith, {\em et~al.} {\em
  Nature} {\bfseries 585} no.~7825, (2020) 357--362.

\bibitem{2020SciPy-NMeth}
P.~Virtanen, R.~Gommers, T.~E. Oliphant, M.~Haberland, T.~Reddy, D.~Cournapeau,
  E.~Burovski, P.~Peterson, W.~Weckesser, J.~Bright, {\em et~al.} {\em Nature
  methods} {\bfseries 17} no.~3, (2020) 261--272.

\bibitem{Clark:2009wm}
M.~A. Clark, R.~Babich, K.~Barros, R.~C. Brower, and C.~Rebbi
  \href{http://dx.doi.org/10.1016/j.cpc.2010.05.002}{{\em Comput. Phys.
  Commun.} {\bfseries 181} (2010) 1517--1528},
  \href{http://arxiv.org/abs/0911.3191}{{\ttfamily arXiv:0911.3191 [hep-lat]}}.

\bibitem{Hunter:2007}
J.~D. Hunter \href{http://dx.doi.org/10.1109/MCSE.2007.55}{{\em Computing in
  Science \& Engineering} {\bfseries 9} no.~3, (2007) 90--95}.

\end{thebibliography}\endgroup

\end{document}